\documentclass[aip,jmp,amsmath,amssymb, reprint]{revtex4-1}
\usepackage{graphicx}
\usepackage{dcolumn}
\usepackage{bm}
\usepackage{amssymb,amsmath}

\begin{document}

\title{$f(R)$ and $f(T)$ theories of modified gravity}


\author{Rafael Ferraro}
\email{ferraro@iafe.uba.ar}
\thanks{Member of Carrera del Investigador Cient\'{\i}fico (CONICET,
Argentina)} \affiliation{Instituto de  Astronom\'\i a y F\'\i sica
del Espacio, Casilla de Correo 67, Sucursal 28, 1428 Buenos Aires,
Argentina, and\\ Departamento de F\'\i sica, Facultad de Ciencias
Exactas y Naturales, Universidad de Buenos Aires, Ciudad
Universitaria, Pabell\'on I, 1428 Buenos Aires, Argentina}

\begin{abstract}
We briefly review $f(R)$ theories, both in the metric and Palatini
formulations, their scalar-tensor representations and the chameleon
mechanism that could explain the absence of perceptible consequences
in the Solar System. We also review $f(T)$ theories, a different
approach to modified gravity consisting in a deformation of the
teleparallel equivalent of General Relativity. We show some
applications to cosmology and cosmic strings. As $f(R)$'s, $f(T)$
theories are not exempted from additional degrees of freedom; we
also discuss this still open issue.
\end{abstract}

\maketitle


\section{Introduction}
In the last five decades many theories of modified gravity have been
proposed in connection with different physical purposes. In the
60's, Brans and Dicke coupled a scalar field $\phi$ to the metric
$g_{\mu\nu}$ to get a variable effective gravitational
constant.\cite{Bra} In Brans-Dicke theory, the scalar field is a new
degree of freedom of the gravitational field, which is not directly
coupled to the matter, but it exerts influence by entering the
dynamical equations that govern the spacetime geometry. The
gravitational Brans-Dicke action contains a new constant $\omega$
that should be dictated by the experiment:
\begin{equation}
S_{BD}[g_{\mu\nu},\phi]=-\frac{1}{2\kappa}\int d^4 x \sqrt{-g}
\left( \phi\, R - \frac{\omega}{\phi}\, g^{\mu\nu}\partial_\mu\phi\,
\partial_\nu\phi\right),
\label{bransdicke}
\end{equation}
where $\kappa\equiv 8\pi G$, and the signature $+---$ was adopted.
Also in the 60's, the Einstein-Hilbert Lagrangian was added with
terms quadratic in the curvature to tackle the renormalization of
the theory.\cite{Uti} In 1971 Lovelock\cite{Lov} considered terms of
higher order in the curvature as well; but he was driven by another
motivation. While this kind of Lagrangians leads, in general, to
fourth order dynamical equations because they contain second order
derivatives, Lovelock obtained the more general Lagrangian polynomic
in the curvature and leading to conserved second order equations for
the metric. Lovelock discovered that the bigger the spacetime
dimension is, the bigger is the number of terms these Lagrangians
can contain. For instance, the Einstein-Lanczos Lagrangian,
\begin{equation}
{L=-\frac{1}{2\kappa}\,(R+2\,\Lambda) + \alpha(R_{\lambda\mu\nu\rho}
R^{\lambda\mu\nu\rho}+R^2-4 R_{\mu\nu}R^{\mu\nu})}\ ,
\end{equation}
is the Lovelock Lagrangian for dimensions 5 or 6; it is the
Einstein-Hilbert Lagrangian with a cosmological constant plus a
quadratic term. If the dimension is 4, then the added quadratic term
becomes a topological invariant (Euler's characteristic); so it does
not contribute to the variation of the action. Then we recover the
Einstein equations as the sole conserved second order equations for
the metric in four dimensions.

In 1970 Buchdahl proposed to replace the Einstein-Hilbert scalar
Lagrangian with a function of the scalar curvature, and studied its
cosmological consequences.\cite{Buc} This type of modified gravity
is nowadays called a $f(R)$ theory. In 1983 Milgrom\cite{Mil}
thought that the galactic rotation curves were an evidence of the
fail of Newtonian gravity to describe gravitation in the weak field
regime ($a_g<<a_o\approx 10^{-10}m s^{-2}$). According to Milgrom,
no dark matter was needed to explain the data but a theory of
modified gravity. In the deep {\it MOND} (Modified Newtonian
Dynamics) regime of Milgrom's theory, the acceleration of gravity
goes to $a_g=\sqrt{a_o\ a_{g_{Newton}}}$. In the last decade
Bekenstein\cite{Bek} developed a relativistic theory of gravity
named {\it TeVeS}, because it combines the metric tensor, a vector
field and a scalar field. TeVeS includes Milgrom's weakening of
Newtonian gravity in the weak field regime, and has also
consequences for lensing phenomena, cosmology, etc.

String theory has been also a source of inspiration for theories of
modified gravity. Just to mention a case, {\it DGP}
gravity\cite{Dva} describes the 4-dimensional universe as immersed
in a 5-dimensional manifold. Thus a ``normal" $5D$ gravity can cause
large scale effects in $4D$, as the accelerated expansion with no
presence of dark energy. These $4D$ consequences are driven by a
scalar field named {\it galileon} because of the symmetries it
obeys.\cite{Nic,Def}

\section{$f(R)$ theories}
The simplest way of modifying Einstein's General Relativity is to
replace the scalar Lagrangian $R$ with a function $f(R)$ of the
scalar curvature:\cite{Buc,Tey,Whi}
\begin{equation}
S\ =\ -\frac{1}{2\kappa} \int d^4 x\ \sqrt{-g} \ f(R)\ .
\label{accion}
\end{equation}
By properly choosing the function $f$, one could generate ``$f(R)$''
theories departing from General Relativity both at small and large
scales. So, deformations at large curvatures could be employed for
smoothing singularities; while deformations at large scales could be
useful to geometrically explain the accelerated expansion without
resorting to dark energy. The weak field regime of the deformed
theory also opens a way to explain phenomena otherwise attributed to
dark matter. References
\onlinecite{Cap11,Sot10,Fel,Tsu,Cap111,Noj06,Noj11} are
comprehensive reviews on $f(R)$ theories.

As it happens in General Relativity, there are two ways of varying a
metric theory of gravity. One can assume the Levi-Civita connection
to write the Ricci tensor; so the metric is left as the sole
dynamical variable. Alternatively, one could regard the affine
connection $\Gamma^\lambda_{\mu\nu}$ and the metric $g_{\mu\nu}$ as
independent dynamical variables. The first way of variation is
called {\it metric formalism}; the second one is the {\it Palatini
formalism}.\cite{Pal} In General Relativity there is no difference
between the results of metric and Palatini formalisms, since the
affine connection making stationary the Einstein-Hilbert action is
precisely the Levi-Civita connection. Actually, the equivalence
between both formalisms is valid for all the Lovelock's
Lagrangians.\cite{Exi} Instead, in $f(R)$ theories both procedures
should be separately studied because they yield different dynamics.
Before choosing one of both formalisms, the variation of the $f(R)$
Lagrangian density can be written as
\begin{eqnarray}
&&\delta\left(f(R) \, \sqrt{-g}\right)\ =\  f'(R)\
\sqrt{-g} \ \delta R\ +\ f(R)\ \delta\sqrt{-g} \nonumber\\
 &&=\   f'(R) \ \sqrt{-g}\ g^{\mu\nu}\ \delta R_{\mu\nu}\label{variacion}\\ &&+\
\sqrt{-g}\, \left(f'(R)\ R_{\mu\nu}-\frac{1}{2}\ f(R) \
g_{\mu\nu}\right)\ \delta g^{\mu\nu}\nonumber\, ,
\end{eqnarray}
where the formula $\delta\ln(\det[g_{\mu\nu}]) = -g_{\mu\nu}\delta
g^{\mu\nu}$, valid for any matrix, was used to vary the determinant
of the metric. Besides, $\delta R_{\mu\nu}$ can be expressed in
terms of variations of the affine connection
$\Gamma^\lambda_{\mu\nu}$ (whatever $\Gamma^\lambda_{\mu\nu}$ is;
see for instance Ref.~\onlinecite{Lan}):
\begin{equation}
\delta R_{\mu\nu}\ =\ \nabla_\lambda\delta\Gamma^\lambda_{\nu\mu}-
\nabla_\nu\delta\Gamma^\lambda_{\lambda\mu}\ .\label{variacionricci}
\end{equation}
Notice that the connection is not a tensor; but the difference
$\delta\Gamma^\lambda_{\mu\nu}$ between two different connections
does transform as a tensor.

\section{Metric formalism for $f(R)$ theories}
If the Levi-Civita connection is assumed, then second derivatives of
$g_{\mu\nu}$ are under variation in Eq.~(\ref{variacionricci}). As a
consequence, fourth-order Euler-Lagrange equations should be
expected as a result of the (double) integration by parts induced by
the variation (\ref{variacionricci}). It is, however, remarkable
that $g^{\mu\nu}\ \delta R_{\mu\nu}$ is, in this case, a
four-divergence. This is because the Levi-Civita connection is {\it
metric}, so the metric can enter the covariant derivative. This is
the reason why General Relativity ($f'(R) = 1$) remains as a theory
governed by second order dynamical equations. Contrarily, $f(R)$
theories in the metric formalism are characterized by fourth-order
dynamical equations:
\begin{equation}
f'(R)\, R_{\mu\nu}-\frac{1}{2}\, f(R)\, g_{\mu\nu} -
\left[\nabla_\mu\nabla_\nu -g_{\mu\nu}\, \Box\right] f'(R)\ =\
\kappa \,T_{\mu\nu}\ .\label{dinamica}
\end{equation}
Notice that $f'(R)$ acts as renormalizing the gravitational constant
$\kappa$; so, only functions with $f'>0$ should be considered
(besides, $f''>0$ to avoid instabilities\cite{Dol,Noj03,Far06}).
Differing from General Relativity, these equations link the scalar
curvature $R$ and the trace $T$ of the energy-momentum tensor not
algebraically but differentially. In fact, the trace of
Eq.~(\ref{dinamica}) is
\begin{equation}
f'(R)\, R-2\, f(R)+3\, \Box f'(R)= \kappa \,T\ ,\label{traza}
\end{equation}
which displays the propagation of a new degree of freedom associated
with $f'(R)$ (this degree of freedom is absent in General Relativity
since it is $f'(R) = 1$).

A $f(R)$ theory can be rephrased as a scalar-tensor theory governed
by second order dynamical equations.\cite{Bar88,Mae,Chi} To show it,
let us start from the following action containing a metric tensor
$g_{\mu\nu}$ and a scalar field  $\phi$:
\begin{equation}
S_{grav}[g_{\mu\nu},\phi]\ =\ -\frac{1}{ 2\kappa }\int d^4 x
\sqrt{-g} \left[ \phi\, R - V(\phi)\right]\ .\label{jordan}
\end{equation}
The variation with respect to $\phi$ gives $R\ =\ V'(\phi)$, so
linking the scalar field to the metric. This result also implies
that the Lagrangian in (\ref{jordan}) is nothing but the Legendre
transform of the function $V(\phi)$; therefore, it depends just on
$R$. So we can call it $f(R)$:
\begin{equation}
f(R)\ \equiv\ \phi\, R - V(\phi)\ .\label{legendre}
\end{equation}
By anti-transforming, one also gets
\begin{equation}
\phi\ =\ f'(R)\ .
\end{equation}
These results show that a $f(R)$ theory in the metric formalism is
dynamically equivalent to the action (\ref{jordan}),  where $f(R)$
and $V(\phi)$ are related through the Legendre transformation
(\ref{legendre}). Notice that $S_{grav}[g,\phi]$ in
Eq.~(\ref{jordan}) resembles a Brans-Dicke theory with $\omega=0$
(absence of kinetic term).

The action (\ref{jordan}) is written in the so called {\it Jordan
frame} representation of the theory. By transforming to the {\it
Einstein frame} representation we will obtain second order dynamical
equations. So let us define
\begin{eqnarray}
\phi\ \rightarrow\ \tilde{\phi} &=& \sqrt{ \frac{3}{2\kappa}}\
\ln\phi\nonumber\\  g_{\mu\nu}\rightarrow \tilde{g}_{\mu\nu} &=&
\phi \, g_{\mu\nu}\, ,\ \ \ \ \ \ \sqrt{-g}\, =\,
\phi^{-2}\sqrt{-\tilde{g}}\, ,\ \ \ \ \ \ \ \label{trans}
\end{eqnarray}
where $\tilde{g}_{\mu\nu}$ is conformally related to ${g}_{\mu\nu}$
through the scalar field $\phi$. Then, one applies the relation for
the scalar curvatures of conformally related metrics,\cite{Wal}
\begin{equation}
\phi\ \tilde{R}\ =\ R - \frac{3}{2}\ g^{\mu\nu}\
\partial_\mu\ln\phi\ \partial_\nu\ln\phi - 3\ \Box\ln\phi\
,\label{conforme}
\end{equation}
and throws out a surface term to write the action in the
form\footnote{Use that $\Box\ln\phi=-g^{\mu\nu}\
\partial_\mu(\ln\phi)\ \partial_\nu(\ln\phi)+\phi^{-1}\Box\phi$, and $\phi^{-2}
\Box=\phi^{-2}(-g)^{-1/2}\partial_\mu ... =
(-\tilde{g})^{-1/2}\partial_\mu ... $.}
\begin{equation}
S'_{grav}[\tilde{g}_{\mu\nu},\tilde{\phi}]\! =\! -\!\!\int d^4x \,
\sqrt{\!-\tilde{g}}  \left[ \frac{\tilde{R}}{2\kappa}-\frac{1}{2}
\tilde{g}^{\mu\nu}\partial_{\mu}\tilde{\phi}\,
\partial_{\nu}\tilde{\phi} + U( \tilde{\phi} ) \right]\label{einstein}
\end{equation}
where the potential is
\begin{equation}
U( \tilde{\phi} )\ =\ -\frac{V(\phi)}{2\kappa \ \phi^2}\ =\
\frac{f(R)-R\, f'(R)}{2\kappa \ f'(R)^2}\ .\label{potencialu}
\end{equation}
As can be seen in Eq.~(\ref{einstein}), the action in the Einstein
frame gets a canonical form: it describes a ``gravitational field''
$\tilde{g}_{\mu\nu}$ and a minimally coupled scalar field
$\tilde{\phi}$, governed by standard second order equations. So, we
have success in reducing the order of the equations (of course, by
widening the number of variables). An easy identification of the
degrees of freedom is now possible: 2 degrees of freedom related to
the tensor $\tilde{g}_{\mu\nu}$ plus 1 degree of freedom associated
with the massive scalar field $\tilde\phi$.

\subsection{The chameleon effect}
Because of the observations in the Solar System, Brans-Dicke theory
is constrained to values $|\omega|>40000$. Despite that metric
$f(R)$ theories have $\omega=0$ in their equivalent representation
(\ref{legendre}), they are not ruled out. This is so because they
contain a potential $V(\phi)$ (or $U(\tilde\phi)$) that is absent in
Brans-Dicke theory. This potential could be useful to hide the
scalar degree of freedom within the Solar System. In other words,
the observations in the Solar System could agree with metric $f(R)$
theories, whenever the scalar degree of freedom does not appreciably
distort the spherically symmetric outer static Schwarzschild
solution for a typical stellar object. Even so, the scalar field
could have physical effects at other scales, as to be the cause of
the accelerated cosmological expansion, etc. This behavior, called
the \emph{chameleon effect}, has been proposed in
Ref.~\onlinecite{Kho04} (cf.
Refs.~\onlinecite{Fel,Olm05,Cem,Fau,Sta07,Nav,Wat}), and is strongly
dependent on the choice of the potential $U$.

The idea can be exemplified by considering solutions to the
dynamical equation (\ref{traza}) for $\phi=f'(R)$. Alternatively,
this equation can be also obtained by adding the action
(\ref{einstein}) with an action for matter minimally coupled to the
metric $g_{\mu\nu}=e^{(-\sqrt{2\kappa/3}\,
\tilde{\phi})}\,\tilde{g}_{\mu\nu}$, and varying with respect to
$\tilde\phi$. We are interested in static spherically symmetric
solutions. In such case, the resulting equation reduces to
\begin{equation}
\nabla^2\tilde{\phi}=\frac{1}{r^2}\,\frac{d}{dr}\!\!\left(r^2\
\frac{d\tilde{\phi}}{dr}\right)= U'(\tilde{\phi})-\rho\
\sqrt{\frac{\kappa}{6}}\ \exp\!\!\left[-\sqrt{\frac{8\kappa}{3}} \
\tilde{\phi}\right], \label{campocamaleon}
\end{equation}
where $\rho\equiv T_{mat}\, ^\mu_\mu >0$. Here we are momentarily
ignoring the back-reaction on the metric, by choosing a Minkowskian
background $\tilde{g}_{\mu\nu}=\eta_{\mu\nu}$.

\begin{figure}
\includegraphics[width=80mm]{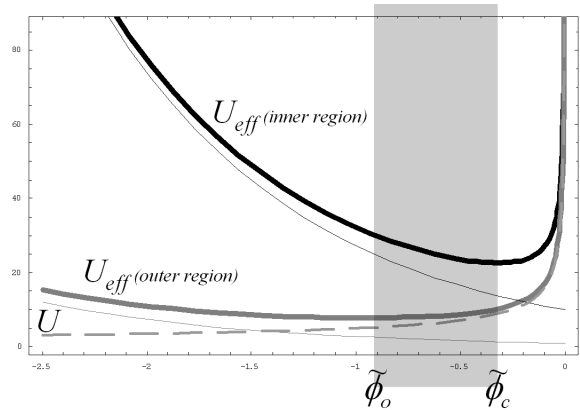}
\caption{Effective potentials in the inner and outer regions. The
(chosen) potential $U$ is the dashed line. The corresponding
exponential terms in Eq.~(\ref{efectivo}) are represented by the
thin lines.} \label{Fig1}
\end{figure}

We will divide the space in two regions of constant density: the
region inner to a spherical star of radius $R_\odot$ and density
$\rho_c$, and the outer region filled by a medium of lower density
$\rho_o$. For a constant energy density $\rho$, we can define (in
each region) the effective potential
\begin{equation}
U_{eff}\equiv U(\tilde{\phi})\ +\ \frac{\rho}{4}\
\exp\left[-\sqrt{\frac{8\kappa}{3}}\,\tilde{\phi}\right]\
.\label{efectivo}
\end{equation}
We have to choose a potential $U$ allowing the chameleon effect to
become apparent. In Figure \ref{Fig1}, $U$ has been chosen in such a
way that $U_{eff}$ has a minimum in both regions (we call them
$\tilde{\phi}_o$ and $\tilde{\phi}_c$). We will search for a
solution varying between $\tilde{\phi}_c$ at the center of the star
and $\tilde{\phi}_o$ at infinity (grey strip in Figure \ref{Fig1}).
Outside the star, we expand $U_{eff}$ at the minimum
$\tilde{\phi}_o$:
\begin{equation}
U_{eff}\simeq m^2(\tilde{\phi}-\tilde{\phi}_o)^2/2+constant \ ,
\end{equation}
where $m$ is the mass of the field in this approach. Thus, the outer
solution is
\begin{equation}
\tilde{\phi}\simeq\tilde{\phi}_o+\frac{C}{r}\, \exp[-m\,
(r-R_\odot)]\ .\label{outer}
\end{equation}
Inside the star, we assume that the exponential term $U_{eff}$
dominates on $U$. Then
\begin{equation}
\nabla^2\tilde{\phi}\ \approx\ -\rho_c\, \sqrt{\frac{\kappa}{6}}\
\exp\left[-\sqrt{\frac{8\kappa}{3}}\,\tilde{\phi}\right]\ \approx\
-\rho_c\, \sqrt{\frac{\kappa}{6}}\ ,
\end{equation}
if $\sqrt{8\kappa/3}\ |\tilde{\phi}|<<1$. So, the inner solution is
\begin{equation}
\tilde{\phi}\ \simeq\ \tilde{\phi}_c\ -\ \rho_c\
\sqrt{\frac{\kappa}{24}}\ R_s^2\,\left(
\frac{r^2}{3\,R_s^2}+\frac{2\,R_s}{3\,r}-1\right)\ .\label{inner}
\end{equation}
Notice that the integration constant $R_s$ fulfills
\begin{equation}
\tilde{\phi}(R_s)=\tilde{\phi}_c\ ,\ \ \ \  \ \ \ \
\tilde{\phi}'(R_s)=0\ .
\end{equation}
So, one can take Eq.~(\ref{inner}) to be the inner solution for
$R_s< r <R_\odot$, and extended it as the constant
$\tilde{\phi}=\tilde{\phi}_c$ to $r=0$ (in fact,
$U'_{eff}(\tilde{\phi}_c)=0$). Thus, $R_s$ in Eq.~(\ref{inner}) and
$C$ in Eq.~(\ref{outer}) remain as two integration constants to be
determined by the continuity of the solution (\ref{inner}) and
(\ref{outer}) and its derivative at $r=R_\odot$. By assuming that
$m\, R_\odot << 1$, one obtains the following two equations for
$R_s$ and $C$:
\begin{equation}
\sqrt{\frac{3}{2\kappa}}\
\left[1-\left(\frac{R_s}{R_\odot}\right)^2\right]\simeq \epsilon\ ,
\end{equation}
\begin{equation}
C\ \simeq\ \sqrt{\frac{2}{3\kappa}}\ \Phi_N\ R_\odot\,
\left[1-\left(\frac{R_s}{R_\odot}\right)^3\right]\ ,
\end{equation}
where $\Phi_N=\kappa\, \rho_c\, R_\odot^2/6$ is the Newtonian
potential on the surface of the star, and
\begin{equation}
\epsilon\equiv \sqrt{\kappa}\ \
\frac{\tilde{\phi}_c-\tilde{\phi}_o}{\Phi_N} \ .\label{epsilon}
\end{equation}
The chameleon effect happens when the potential $U$ is such that
$\epsilon << 1$. In fact, in such case it is
\begin{equation}
\sqrt{\kappa}\ C\ \approx\ \epsilon\ \Phi_N\ R_\odot\ <<\ \kappa\,
M_\odot\ ,\label{camaleon2}
\end{equation}
so the effect of the potential $\tilde\phi$ around the star is
negligible compared with Newtonian gravity.\footnote{Notice that
$g_{00}\simeq 1-\sqrt{2\kappa/3}\,\tilde\phi$, because
$\sqrt{\kappa}\,|\tilde\phi|<<1$.  Therefore $g_{00}$ implies a weak
Yukawa-shaped gravitational potential
$\Phi(r)=-\sqrt{\kappa/6}\,\tilde\phi(r)$ with $m\, R_\odot<<1$.
Because of Eq.~(\ref{camaleon2}), $\Phi$ can be ignored, which is
the realization of the chameleon effect.} Besides,
\begin{equation}
\frac{R_\odot-R_s}{R_\odot}\ \approx\ \frac{\epsilon}{\sqrt{6}}\ <<\
1\ ,
\end{equation}
then the inner solution differs from $\tilde{\phi}_c$ just in a
thin-shell near the surface.

The back-reaction on the metric has been considered in
Ref.~\onlinecite{Fau}; it is proved that the PPN parameter
characterizing the departure from the Schwarzschild metric is
$\gamma\ \simeq\ 1\, +\, \sqrt{2/3}\ \epsilon$. The Cassini tracking
constrains $\epsilon$ to be $\epsilon\lesssim 10^{-5}$ in the Solar
System. Since the Newtonian potential on the Sun surface is
$\Phi_{Sun}\sim 10^{-6}$, one obtains
\begin{equation}
\sqrt{\kappa}\ (\tilde{\phi}_{Sun}-\tilde{\phi}_o)\ \lesssim\
10^{-11}\ .\label{cota}
\end{equation}
The viable $f(R)$ theories are those having a potential $U$
accomplishing this relation. A typical $f(R)$ used to model the
accelerated expansion is:\cite{Fau,Nav,Carr,Ame}
\begin{equation}
f(R)\ =\ R\ +\ \frac{\mu^{2(n + 1)}}{R^n}\ ,\label{acelerada}
\end{equation}
because the $R^{-n}$ term dominates at low curvature. By replacing
the potential $U$ of Eq.~(\ref{potencialu}) in Eq.~(\ref{efectivo}),
one gets
\begin{equation}
U_{eff}(\tilde\phi)\ =\ -\frac{(n+1)\, \mu^2}{2\kappa\,
\phi^2}\left(\frac{1-\phi}{n}\right)^\frac{n}{n+1}+\frac{\rho}{4\,
\phi^2}\ ,\label{ueff}
\end{equation}
where $\phi = \exp[\sqrt{2\kappa/3} \, \tilde\phi]$. If
$\rho\kappa>>\mu^2$, the minima of $U_{eff}$ in each region are
\begin{equation}
\sqrt{2\kappa/3}\ \tilde{\phi}_{o,c}\simeq
-n\left(\frac{\mu^2}{\rho_{o,c}\kappa}\right)^{n+1} \
,\label{minimo}
\end{equation}
which are very near to zero as required by Eq.~(\ref{cota}). This
also means that there is no sensitive difference between metrics
$g_{\mu\nu}$ and $\tilde{g}_{\mu\nu}$ (see Eq.~(\ref{trans})). In
the Solar System, the density $\rho_o$ outside the Sun should be
replaced by the mean density of the barionic matter in the galaxy
($\rho_{galaxy}\kappa$ is around $10^5$ times the squared Hubble
constant). Since $\rho_{galaxy}<<\rho_{Sun\ }$, the Eq.~(\ref{cota})
becomes
\begin{equation}
n\left(\frac{\mu^2}{\rho_{galaxy}\kappa}\right)^{n+1}\ \lesssim\
10^{-11}\ ,
\end{equation}
in order that the model (\ref{acelerada}) be acceptable. The above
used condition $m R_\odot<<1$ can be accomplished too; in fact
\begin{equation}
m^2\ =\ U''_{eff}(\tilde{\phi}_o)\ \sim \rho_{galaxy}\kappa\,
\left(\frac{\rho_{galaxy}\kappa}{\mu^2}\right)^{n+1}\, .
\end{equation}

\subsection{Metric $f(R)$ theories in cosmology}
\begin{figure}
\includegraphics[width=60mm]{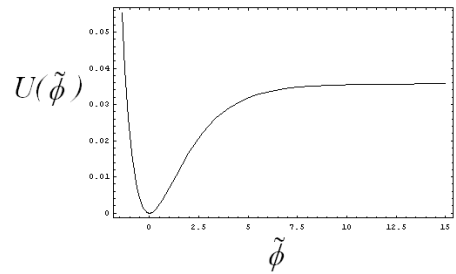}
\caption{Potential $U(\tilde\phi)$ in the model $f(R)\ =\ R\ +\
\alpha\ R^2$.}\label{Fig2}
\end{figure}
Within the framework of a FRW universe, it has been shown that the
field $\tilde\phi$ is attracted to the minimum of the potential
(\ref{ueff}), and then adiabatically evolves following the
Eq.~(\ref{minimo}) with $\rho=\rho_{universe}(t)$. Using the usual
approximations, it is obtained that the net effect of the presence
of $\tilde\phi$ is the adding of a constant to the density of
matter.\cite{Fau,Brax} In such case, the cosmological effects
resulting from the model (\ref{acelerada}) would be
undistinguishable from a mere cosmological constant (``vanilla dark
energy''). The growing of inhomogeneities are, however, a more
promising arena to distinguish among $f(R)$ theories and the
$\Lambda CDM$ model.\cite{Son,Car,Cru}

$f(R)$ theories have also be applied to modify the high curvature
regime. The simplest example is
\begin{equation}
f(R)\ =\ R\ +\ \alpha\ R^2\ ,
\end{equation}
which produces inflation, with $\tilde\phi$ playing the role of
inflaton.\cite{Sta80,Hwa} Figure \ref{Fig2} shows that the potential
$U(\tilde\phi)$ is nearly flat for large values of $\tilde\phi$, as
required to get inflation.

Other cosmological effects, such as lensing due to overdensities of
matter, have also been considered in the framework of metric $f(R)$
theories.\cite{Zha}

\section{Palatini formalism for $f(R)$ theories}
In Palatini formalism\cite{Pal,Ein} the connection
$\Gamma^\lambda_{\mu\nu}$ and the metric $g_{\mu\nu}$ are regarded
as dynamical variables to be independently varied. Thus, $\nabla$ in
Eq.~(\ref{variacionricci}) is just the covariant derivative for an
arbitrary connection $\Gamma^\lambda_{\mu\nu}$. The variation of the
action with respect to the connection involves the integration by
parts of the first term in the Eq.~(\ref{variacion}); what results
is not a dynamical equation but a constraint for the
connection:\cite{Ham,Vol03,Olm05}
\begin{eqnarray}
\Gamma^\lambda_{\phantom{a}\mu\nu}\ =\ \frac{g^{\lambda\sigma}}{2\,
f'(R)}\, \Big[&&\partial_\mu \left(f'(R)\, g_{\nu\sigma}\right)\
+\ \partial_\nu \left(f'(R)\, g_{\mu\sigma}\right)\nonumber\\
&&-\ \partial_\sigma \left(f'(R)\, g_{\mu\nu}\right)\Big]\
\label{conexionpalatini}
\end{eqnarray}
(this is the result when torsion is neglected\cite{Olm11,Ort}).
General Relativity is a special case, in the sense that the
connection (\ref{conexionpalatini}) is the Levi-Civita connection
when $f(R)=R$; so, no difference exists between metric and Palatini
formalisms in General Relativity. But, in a general case, the
connection (\ref{conexionpalatini}) is not metric for $g_{\mu\nu}$,
but for the conformal metric $\tilde{g}_{\mu\nu}\, =\, f'(R)\
g_{\mu\nu}$.

On the other hand, the variation of the action with respect to the
metric in Eq.~(\ref{variacion}) does yield dynamical equations:
\begin{equation}
f'(R)\, R_{\mu\nu} - \frac{1}{2}\, f(R)\, g_{\mu\nu} = \kappa \,
T_{\mu\nu}\ ,\label{palatinidinamica}
\end{equation}
where $R_{\mu\nu}$ and $R$ are built with the connection
(\ref{conexionpalatini}). Here $T_{\mu\nu}$ is the usual
energy-momentum tensor, whenever the action for matter does not
contain covariant derivatives. As a remarkable difference compared
with the metric formalism, the trace of Eq.~(\ref{palatinidinamica})
does not govern the propagation of a scalar degree of freedom but it
is a mere algebraic relation between the curvature $R$ and the
matter distribution:
\begin{equation}
f'(R)\, R\ -\ 2\, f(R)\ =\ \kappa \, T\ ,\label{algebraica}
\end{equation}
so suggesting that Palatini formalism does not harbor additional
degrees of freedom, as confirmed by means of the Einstein frame
representation of Palatini dynamics.\cite{Sot10} However, the
relation $R=R(T)$ in Eq.~(\ref{algebraica}) implies that the
connection (\ref{conexionpalatini}) depends on first derivatives of
$T$. Therefore, Eq.~(\ref{palatinidinamica}) involves second
derivatives of $T$, which is a very unlike coupling between geometry
and matter. This feature is the source of several troubles: juncture
conditions on the surface of spherically symmetric bodies leading to
curvature divergences even for reasonable state equations of
matter\cite{Bar08,Bar081} (however, see Ref.~\onlinecite{Olm081}),
incompatibilities with the stability of microscopic
systems\cite{Vol05,Fla,Olm08} and non-well formulated Cauchy problem
unless the trace $T$ is constant.\cite{Far}


\section{$f(T)$ theories}
General Relativity can be reformulated in a {\it teleparallel}
framework by taking the field of orthonormal frames or
\emph{tetrads} as the dynamical variable instead of the metric
tensor.\cite{Ein30} The tetrad is a basis $\{{\bf e}_a({\bf x})\}$,
$a = 0,1,2,3$, of vectors in the spacetime. Each vector ${\bf e}_a$
can be decomposed in a coordinate basis, so giving the components
$e_a^\mu$; thus, the orthonormality condition reads:
\begin{equation}
\eta_{ab}\ =\ g_{\mu\nu}\, e_a^\mu\, e_b^\nu\ ,\label{ortonormal}
\end{equation}
where $\eta_{ab}=\textrm{diag}(1,-1,-1,-1)$. This relation can be
inverted with the help of the co-frame $\{{\bf e}^a\}$, defined as
\begin{equation}
e_a^\mu\ e^b_\mu\ =\ \delta_a^b\ ,
\end{equation}
to obtain the metric starting from the tetrad:
\begin{equation}
g_{\mu\nu}\ =\ \eta_{ab}\, \, e^a_\mu\, e^b_\nu\ \ \ \Rightarrow\ \
\ \sqrt{-g}\ =\ \det[e^a_\mu]\ \equiv\ e\ .\label{metrica}
\end{equation}

The Teleparallel Equivalent of General Relativity (TEGR) is a theory
for the tetrad, whose dynamical equations are equivalent to Einstein
equations whenever the tetrad is related to the metric through the
Eq.~(\ref{metrica}). The TEGR Lagrangian does not contains second
derivatives because it is quadratic in the tensor
\begin{equation}
T^\mu_{\ \ \nu\rho}\ =\ e^\mu_a\ (\partial_\nu
e_\rho^a-\partial_\rho e_\nu^a)\ ,\label{torsion}
\end{equation}
which is reminiscent of the electromagnetic field tensor (in fact,
it is built of the set of four exact 2-forms $T^a\equiv d{\bf
e}^a$). The tensor (\ref{torsion}) can be regarded as the
\emph{torsion} of the Weitzenb\"{o}ck connection,
\begin{equation}
\Gamma\ ^\mu_{\rho\nu}\ \equiv\ e_a^\mu\ \partial_\nu e_\rho^a\ =\
-e^a_\rho\ \partial_\nu e_a^\mu\, .\label{Weitzenbock}
\end{equation}
Weitzenb\"{o}ck spacetime has torsion but it is flat, because the
Riemann tensor associated with the connection (\ref{Weitzenbock}) is
identically null. The connection (\ref{Weitzenbock}) has the nice
property that a vector is parallel-transported iff its projections
on the tetrad remain constant; in fact, $\nabla_\nu V^\mu =
e_a^\mu\,
\partial_\nu(e^a_\lambda\, V^\lambda)$. Moreover, Weitzenb\"{o}ck
connection is metric compatible since $\nabla_\nu e^\mu_a\equiv 0$.
Weitzenb\"{o}ck connection could be compared with Levi-Civita
connection by using Eq.~(\ref{metrica}). It results that they differ
in a tensor named \emph{contorsion}. The contorsion takes part in
the TEGR Lagrangian, since the TEGR action is:\cite{Hay,Arc}
\begin{equation}
S_T[\textbf{e}^a]\ =\ \frac{1}{2\kappa}\ \int d^4x\,\ e\,\ S_\rho^{\
\ \mu\nu}\, T^\rho_{\ \ \mu\nu} \ \equiv\ \frac{1}{2\kappa}\ \int
d^4x\,\ e\,\ \mathbb{S}\cdot \mathbb{T}\ ,
\end{equation}
where
\begin{equation}
2\, S_\rho^{\ \ \mu\nu} \equiv \underbrace{\frac{1}{2}\, (T_\rho^{\
\, \mu\nu}-T^{\mu\nu}_{\ \ \, \, \rho}+T^{\nu\mu}_{\ \ \, \,
\rho})}_{\textrm{\it contorsion}\ K^{\mu\nu}_{\ \ \, \rho}}\, +\,
T^{\ \, \lambda\mu}_\lambda \ \delta^\nu_\rho\,-\, T^{\ \,
\lambda\nu}_\lambda \ \delta^\mu_\rho\, .
\end{equation}
The equivalence between TEGR action and Einstein-Hilbert action
comes from the fact that their Lagrangians differ in a
four-divergence:
\begin{equation}
-e\ R[e^a]\ =\ e\ \, \mathbb{S}\cdot \mathbb{T}\ -\ 2\
\partial_\rho(e\,T^{\mu\ \ \rho}_{\ \ \mu})\ ,\label{equivalencia}
\end{equation}
where $R[e^a]$ is the scalar curvature for the Levi-Civita
connection, with the metric replaced with (\ref{metrica}). In
particular, the four-divergence encapsulates all the second
derivatives contained in the Einstein-Hilbert Lagrangian.

In the same spirit than a $f(R)$ theory, a $f(T)$ theory consists in
a deformation of the TEGR Lagrangian:\cite{Fer07,Fer08,Ben,Lin,Li}
\begin{equation}
S_T\ =\ \frac{1}{2\kappa}\ \int d^4x\ \, e\, \ \mathbb{S}\cdot
\mathbb{T}\ \longrightarrow\ S\ =\ \frac{1}{2\kappa}\ \int d^4x\ \,
e\, \ f(\mathbb{S}\cdot \mathbb{T})\ .
\end{equation}
But, differing from $f(R)$ theories, the dynamical equations in
$f(T)$ theories are always second order because the Lagrangian does
not contain second derivatives. For matter coupled to the metric in
the usual way, they are
\begin{eqnarray}
\nonumber &&4\ \left[e^{-1}\ \partial_\mu(e\ S_a^{\ \ \mu\nu})\ +\
e_a^\lambda
\ T^\rho_{\ \ \mu\lambda}\ S_\rho^{\ \ \mu\nu}\right]\ f'(\mathbb{S}\cdot \mathbb{T})\\
&&+\ 4\ S_a^{\ \ \mu\nu}\ \partial_\mu(\mathbb{S}\cdot \mathbb{T})\
f''(\mathbb{S}\cdot \mathbb{T})\ -\ e_a^\nu \ f(\mathbb{S}\cdot
\mathbb{T})\nonumber\\ &&\ \ \ \ \ \ \ \ \ \ \ \ \ \ \ \ \ \ \ \ \ \
\ \ \ \  =\ -2\kappa\ e_a^\lambda\ T_\lambda^{\ \nu}\ ,
\label{ecuaciones}
\end{eqnarray}
where $T_\lambda^{\ \nu}$ is the energy-momentum tensor.

\subsection{Cosmology}
The first $f(T)$ model was proposed to avoid the Big-Bang
singularity and obtain inflation without resorting to an
inflaton.\cite{Fer07} But most of the cosmological applications
concentrated in the late accelerated expansion of the
universe.\cite{Ben,Lin,Wu10,Kar,Myr,Tsy,Bam} A flat FRW universe is
described by
\begin{eqnarray}
&&e^a_\mu\ =\ \textrm{diag} [1,a(t),a(t),a(t)]\ \textrm{in
comoving coordinates,}\nonumber\\
&&\mathbb{S}\cdot \mathbb{T}\ =\ -6\ H^2\ , \label{flatFRW}
\end{eqnarray}
where $H\, \equiv\,\dot{a}/a$ is the Hubble parameter. Thus, the
dynamical equations (\ref{ecuaciones}) become
\begin{eqnarray}
12\ H^2\ f'(-6 H^2)\ +\ f(-6 H^2)\ &=&\ 2\kappa\ \rho\ ,\label{FRW}\\
-4\, {H\!\!\!\!^{^{^{\centerdot}}}}\ f'(-6 H^2)\,+\,48\,H^2\,
{H\!\!\!\!^{^{^{\centerdot}}}}\ f''(-6 H^2)\ &=&\ 2\kappa\ (\rho+p)\
,\nonumber
\end{eqnarray}
where $\rho$ and $p$ are the energy density and pressure of the
fluid of matter (the conservation law $\dot\rho=-3H(\rho+p)$ is
guaranteed by Eq.~(\ref{FRW})).

In Ref.~\onlinecite{Fer07} a high curvature deformation,
$f(T)=\lambda(\sqrt{1+2T/\lambda}-1)$, was proposed to correct the
evolution near the Big-Bang (more precisely, when $|T|$ is of the
order of $\lambda$). It was found that the Big-Bang is removed and
replaced with an exponential expansion ($H(t)$ goes to
$\sqrt{\lambda/12}$ when $t\rightarrow -\infty$) for any state
equation $p=w\,\rho$ with $w>-1$. As a consequence, the particle
horizon diverges and the whole universe turns out to be causally
connected.

Other no less important issues, such as the growth of
fluctuations,\cite{Den,Che,Zhe} the observational
constraints\cite{Wu101,Wu102,Ben1} or the variation of the universal
constants,\cite{We11,We111} have also been studied in $f(T)$
cosmology.

\subsection{Cosmic strings}
Static circular\cite{Gon11} or spherically\cite{Wa11,Boe} symmetric
solutions are also analyzed in the $f(T)$ literature. In particular,
it has been shown that the Schwarzschild geometry remains as a
solution of $f(T)$ theories.\cite{Fer11} The issue of removing
singularities in stationary configurations was studied in a slightly
different framework of modified TEGR, by using a Lagrangian density
inspired in Born-Infeld electrodynamics:\cite{Fer10}
\begin{eqnarray}
\mathcal{L}\ =\ -\frac{\lambda}{2\kappa}\ &&\Big[\sqrt{\det[g_{\mu
\nu
}-2\lambda ^{-1}F_{\mu \nu}]}-\sqrt{-g}\Big]\nonumber\\
\xrightarrow[\lambda\rightarrow\infty]\ \ &&\frac{1}{2\kappa}\
\sqrt{-g}\ Tr(F)\ ,\label{determinantal}
\end{eqnarray}
($g_{\mu\nu}$ is that of Eq.~(\ref{metrica})). TEGR is recovered in
the limit $\lambda\rightarrow\infty$ if $Tr(F)=\mathbb{S}\cdot
\mathbb{T}$. A possible choice, but not the only one, is $F_{\mu \nu
} = S_{\mu \lambda \rho }\, T_{\nu }^{\,\,\,\lambda \rho }$; then
\begin{equation}
\mathcal{L} \ =\ \frac{e}{2\kappa}\ \left[ \mathbb{S}\cdot
\mathbb{T}\, -\, \frac{\lambda^{-1}}{2}(\mathbb{S}\cdot
\mathbb{T})^{2} \, +\, \lambda^{-1}\, F_{\mu }^{\,\,\nu }F_{\nu
}^{\,\,\mu }\right] +\mathcal{O}(\lambda ^{-2})\,
.\label{desarrollo}
\end{equation}
This expression shows that the theory (\ref{determinantal}) modifies
General Relativity at high curvatures and differs from a mere $f(T)$
theory. These features make it potentially able of avoiding the
singular Schwarzschild solution or any other solution having
$\mathbb{S}\cdot \mathbb{T}=0$.\cite{Fer08,Fer11}

The Lagrangian (\ref{determinantal}) was used in
Ref.~\onlinecite{Fer10} to heal the singular behavior of a cosmic
string:
\begin{equation}
ds^2\ =\ d(t+4 J\,\theta)^2 - Y^2(\rho)\, d\rho^2 - \rho^2\, M^2\,
d\theta^2 - dz^2\, .\label{string}
\end{equation}
In General Relativity it is $Y=1$. In particular, if the dimension
is reduced to $D=2+1$ ($z$ is removed), the $Y=1$ case is a solution
of Einstein equations for $T^{00}=\mu\, \delta(x,y)$ and
$T^{0i}=(J/2)\, \epsilon^{ij}\,
\partial_j \delta(x,y)$, where $\mu\equiv (1-M)/4$. So the solution
(\ref{string}) looks like the geometry associated with a particle of
mass $\mu$ and spin $J$ (a \emph{cosmon}).\cite{Des} However, no
gravitational field surrounds the cosmon since the metric is
manifestly flat (in terms of the Levi-Civita curvature). Instead,
the presence of a cosmon only produces topological effects: the
deficit angle $8\pi\mu$ (conical singularity), and the existence of
closed timelike curves (CTC) of constant $(t, \rho, z)$ when
$\rho<\rho_o\equiv 4J/M$:
\begin{equation}
ds^2=-\left(\rho^2-\frac{16 J^2}{M^2}\right)\, M^2\, d\theta^2\, .
\end{equation}
When the geometry (\ref{string}) is treated within the modified
gravity framework ruled by the Lagrangian (\ref{determinantal}),
then $Y$ becomes a $J$-depending function of $\rho$. $Y(\rho)$ goes
to $1$ for $\rho>>4J/M$ (GR limit) but diverges for $\rho\rightarrow
4J/M$.\cite{Fer10,Fer112} Besides, the solution $J=0$ coincides with
the respective GR solution. While $J$ in General Relativity has no
local effects ($J$ could be locally absorbed through the coordinate
change $t'=t+4 J\,\theta$), now the integration constant $J$ is a
physically relevant degree of freedom that fixes the scale ruling
the GR limit. The curved geometry that replaces the GR cosmic string
has remarkable features: i) the Levi-Civita curvature is well
behaved at $\rho_o = 4J/M$ ($R$, $R_{\mu\nu}R^{\mu\nu}$ and
$R_{\lambda\mu\nu\rho}R^{\lambda\mu\nu\rho}$ vanish at $\rho_o$),
ii) an infinite proper time is required to reach $\rho_o$, and iii)
no CTC's are left. So, the theory (\ref{determinantal}) successfully
smoothes the GR cosmic string.


\subsection{Degrees of freedom in $f(T)$ theories}

$f(T)$ gravity is structurally simpler than metric $f(R)$ theories,
because it always produces second order dynamical equations.
However, this nice feature does not prevent $f(T)$ gravity from
displaying additional degrees of freedom. The circular symmetric
solution of the previous section, obtained in the context of an
extension of $f(T)$ gravity, exhibits a local degree of freedom
associated with the integration constant $J$ that is only globally
apparent in the corresponding GR solution. Since $f(T)$ is a theory
not for the metric but for the tetrad, it contains more degrees of
freedom than GR. In fact, many different tetrads lead to the same
metric, since the relation (\ref{metrica}) is invariant under local
Lorentz transformation of the tetrad field. In order that a theory
for tetrads has the same degrees of freedom than a theory for the
metric, its action should be invariant under local Lorentz
transformation of tetrads in the tangent space. TEGR is a particular
case accomplishing this condition: although $\mathbb{S}\cdot
\mathbb{T}$ does vary under local Lorentz transformation of the
tetrad field, the variation is located in the divergence term of
Eq.~(\ref{equivalencia}); therefore the dynamics does not vary. But
in a $f(T)$ theory, the variation affects the dynamics because the
divergence term remains encapsulated in the function $f$. Only a
global Lorentz invariance survives in such case. Because of this
reason, a $f(T)$ theory globally determines the field of tetrads; it
provides the spacetime with a global frame that fixes its metric and
endows it with a \emph{parallelization}. A local Lorentz
transformation would destroy the parallelization (consider, for
instance, a Cartesian grid in Minkowski spacetime). As was proven in
Ref.~\onlinecite{Sot11}, the local Lorentz invariance cannot be
restored by adding the action with a spin connection. The issue of
counting the number of degrees of freedom in a $f(T)$ theory could
be tackled by reformulating the $f(T)$ action in a Brans-Dicke-like
form.\cite{Yan} Nevertheless, the counting the first and second
class constraints in the Hamiltonian formulation shows that $f(T)$
theories in four dimensions have five degrees of freedom.\cite{Li1}

Summarizing, in passing from TEGR to $f(T)$ gravity, we are
replacing a local symmetry with a global one; in return, we would be
converting global degrees of freedom (like the topological $J$ in
the cosmic string) into local degrees of freedom. These new local
degrees of freedom could be essential to heal singularities. As a
last remark, it should be realized that, even if the geometry is
highly symmetric, it could be very hard to exploit the symmetry to
anticipate aspects of the tetrad field parallelizing such a
geometry. This causes that the naive diagonal choice we used in
Eq.~(\ref{flatFRW}) does not work for open and closed FRW
universes.\cite{Fer111}

\section{Conclusions}
$f(R)$ and $f(T)$ theories are alternative ways to modify General
Relativity. Like metric $f(R)$ theories, $f(T)$ gravity contains
additional degrees of freedom. However, these additional degrees of
freedom do not appear as a consequence of the higher order of the
dynamical equations, since $f(T)$ gravity always leads to second
order equations. They appear because $f(T)$ gravity provides the
spacetime not only with a metric but with a global parallelization.
An extension of $f(T)$ gravity --the one governed by the
determinantal Lagrangian of Eq.~(\ref{determinantal})-- shows that
the extra degrees of freedom can play a fundamental role in
smoothing singularities. In the case of the cosmic string, a global
(topological) property of the GR solution, encoded in the constant
$J$, becomes a local degree of freedom entering the metric tensor.
This generates a family of geometries parametrized by $J$, which
includes a GR solution as a particular case (the $J=0$ case).

\begin{acknowledgments}
The author wishes to thank the organizers of I CosmoSul. This work
was supported by Consejo Nacional de Investigaciones
Cient\'{\i}ficas y T\'{e}cnicas (CONICET) and Universidad de Buenos
Aires.
\end{acknowledgments}



\end{document}